
\magnification=1200

\def\bs{\bigskip}

\def\nt{\noindent}

\def\ce{\centerline}
\def\eq{\eqno}

\def\section#1#2{\vskip1truecm\nt{\bf#1\ #2}\vskip0.5truecm}\indent
\def\subsection#1#2{\vskip1truecm\nt{\bf#1\ #2}\vskip0.5truecm}\indent
\def\subsubsection#1#2{\vskip1truecm\nt{\bf#1\ #2}\vskip0.5truecm}\indent
\indent


\def\G{\Gamma}
\def\d{\delta}

\def\q{\theta}

\def\t{\tau}
\def\f{\phi}

\def\vf{\varphi}
\def\w{\omega}



\def\o{\over}
\def\p{\partial}
\def\ra{\rightarrow}

\def\im{\imath}


\def\sqr#1#2{{\vcenter{\vbox{\hrule height.#2pt\hbox{\vrule width.#2pt
height#1pt \kern#1pt\vrule width.#2pt}\hrule height.#2pt}}}}


\def\boxit#1{\vbox{\hrule\hbox{\vrule\kern3pt
\vbox{\kern3pt#1\kern3pt}\kern3pt\vrule}\hrule}}

\def\dq{{\dot q}}

\ce{\bf QUANTUM MECHANICS OF TIME-DEPENDENT SYSTEMS}

\bs

\ce{\bf CONSTRUCTION OF PURE STATES}

\bs\bs\bs

\ce{{\bf Victor Tapia}\footnote{*}{e-mail: VTAPIA@HALCON.DPI.UDEC.CL}}

\bs

\ce{Departamento de F{\'\i}sica}
\ce{Facultad de Ciencias F{\'\i}sicas y Matem\'aticas}
\ce{Universidad de Concepci\'on}
\ce{Casilla 4009, Concepci\'on, Chile}

\bs\bs

\nt Short title: QUANTUM MECHANICS OF TIME-DEPENDENT SYSTEMS

\bs\bs

\nt PACS: 03.65-w Quantum theory; quantum mechanics

\bs\bs

{\bf Abstract.} For time-dependent systems the wavefunction depends
explicitly on time and it is not a pure state of the Hamiltonian. We
construct operators for which the above wavefunction is a pure state. The
method is based on the introduction of conserved quantities $Q$ and the pure
states are defined by ${\hat Q}\psi=q\psi$. The conserved quantities are
constructed using parametrised mechanics and the Noether theorem.

\bs\bs
\nt{\bf 1. Introduction}
\bs

The quantum mechanics of time-dependent systems has been the subject of
several investigations and in spite of the many efforts done on the subject
it seems that still there is not even agreement on what {\it quantum
mechanics of time-dependent systems} means.

There are several ways in which one can abord the quantum mechanics of
time-dependent systems. For our purposes we will restrict our considerations
to systems which can be described by means of a phenomenological effective
Lagrangian, $L(q,\dq,t)$, or Hamiltonian, $H(q,p,t)$. In this case one can
directly write down a Schr\"odinger equation

$${\hat H}\,\psi=H\left(q,\,-\im\,\hbar\,{\p\o{\p q}},\,t\right)\,\psi=\im\,
\hbar\,{{\p\psi}\o{\p t}}\,.\eq{(1.1)}$$

\nt The wave function, $\psi(q,t)$, is explicitly time-dependent and
therefore one looses the familiar concept of stationary states. Furthermore,
one looses the concept of pure states (those states having a well-defined
quantum number with respect to a given observable) and it is not clear if
there exist other observables with respect to which the wavefunction is a
pure state.

This situation can however be corrected if we are able to find conserved
quantities $Q$ for the time-dependent system. In this case one can implement
the quantum conditions

$${\hat Q}\,\psi=q\,\psi\,.\eq{(1.2)}$$

\nt Therefore, even when the wave function is time-dependent, there would be
observables with respect to which the wavefunction behaves as a pure state.
The problem is reduced therefore to the construction of conserved quantities
for time-dependent systems.

In this work we introduce a novel method for the construction of conserved
quantities for time-dependent systems based on parametrised mechanics. In
this approach the time $t$ is treated as a configuration variable. The system
therefore looks as a time-independent system and conserved quantities can be
found by means, in our case for example, of the Noether theorem.

Let us now describe our method more in detail. First of all we will restrict
our considerations to systems for which there exist a description in terms of
a time-dependent effective Lagrangian, $L(q,\dq,t)$. This restriction is in
view of the clear prescription for quantisation which exist in this case.

For time-independent Lagrangians, $L(q,\dq)$, the time $t$ can be considered,
in some suitable sense, as an ignorable variable. In fact, the Hamiltonian is
time-independent, $H(q,p)$, and is furthermore a conserved quantity,
$dH/dt=0$. Furthermore the Schr\"odinger equation can be solved in terms of a
complete set of time-independent wavefunctions $\psi_n(q)$. This leads
furthermore to the concept of pure states.

The situation is quite different for time-dependent Lagrangians,
$L(q,\dq,t)$. In fact, the Hamiltonian is time-dependent, $H(q,p,t)$, and is
no more a conserved quantity, $dH/dt\not=0$. Furthermore, the solution to
the Schr\"odinger equation is an explicitly time-dependent wavefunction
$\psi(q,t)$. Finally, one looses the concept of pure state.

Therefore, for time-dependent Lagrangians the time $t$ can no more be ignored
so easily, it is no more an ignorable variable. However, it is neither a
variable at the same level than the usual configuration variables $q$. The
ideal situation would be to have the time $t$ at the same level than
configuration variables $q$, since in that case (only configuration variables
and an ``ignorable'' time) we would be able to construct pure states.
Therefore, we must raise the time $t$ to the level of a configuration
variable, $t\ra q^0$. This is achieved by introducing a new variable $\t$
playing the role of the old time. In this case the Lagrangian becomes
homogeneous at the first-order in the derivatives with respect to the new
time variable

$${\bar L}=L\left(q^i,\,{{\dq^i}\o{\dq^0}},\,q^0\right)\,\dq^0\,.\eq{(1.3)}$$

\nt This corresponds to a particular kind of constrained systems. One can
check that the Hamiltonian is identically zero, ${\bar H}\equiv0$.
Furthermore there exists a first-class constraint

$$\f={\bar p}_0+H(q^i,\,q^0,\,{\bar p}_i)\approx0\,.\eq{(1.4)}$$

\nt For constrained systems quantum mechanics is done in terms of first-class
constraints, ${\hat\f}\psi=0$, which in our case reduces to

$$-\im\,\hbar\,{{\p\psi}\o{\p t}}+H\left(q^i,\,t,\,-\im\,\hbar\,{\p\o{\p q^i}
}\right)\,\psi=0\,,\eq{(1.5)}$$

\nt which we recognise as the original Schr\"odinger equation. Therefore the
Lagrangian ${\bar L}$ is dynamically equivalent to the original $L$.

The conserved quantities associated to ${\bar L}$ are constructed with the
Noether theorem. We introduce the Noether variation of the Lagrangian

$$\xi({\bar L})=\xi^0\,{{\p{\bar L}}\o{\p q^0}}+\xi^i\,{{\p{\bar L}}\o{\p q^i
}}+{\dot\xi}^0\,{{\p{\bar L}}\o{\p\dq^0}}+{\dot\xi}^i\,{{\p{\bar L}}\o{\p\dq^
i}}\,.\eq{(1.6)}$$

\nt If $\xi({\bar L})=0$ then

$$Q(q^i,\,q^0,\,{\bar p}_i)=-\xi^0(q^i,\,q^0)\,H(q^i,\,q^0,\,{\bar p}_i)+\xi^
i(q^i,\,q^0)\,{\bar p}_i\,.\eq{(1.7)}$$

\nt is a conserved quantity. Now we can use the conserved quantity above to
obtain pure states. The corresponding equation is

$${\hat Q}\,\psi=\left[-\xi^0(q^i,\,q^0)\,H(q^i,\,q^0,\,{\hat p}_i)+\xi^i(q^i
,\,q^0)\,{\hat p}_i\right]\,\psi=q\,\psi\,.\eq{(1.8)}$$

As an example we consider the damped harmonic oscillator described by the
equation

$${\ddot x}+2\,\G\,{\dot x}+{\w_0}^2\,x=0\,,\eq{(1.9)}$$

\nt The corresponding effective Lagrangian is

$$L={m\o2}\,({\dot x}^2-{\w_0}^2\,x^2)\,e^{2\G t}\,.\eq{(1.10)}$$

\nt The corresponding wave functions are

$$\eqalignno{\psi_n(x,\,t)&=N_n\,H_n\left(\left({{m\w}\o\hbar}\right)^{1/2}\,
e^{\G t}\,x\right)\cr
&\quad\times\exp\left(-{m\o{2\hbar}}(\w+\im\G)e^{2\G t}x^2\right)\,\exp\left(
-\im\left(n+{1\o2}\right)\w t+{\G\o2}\,t\right)\,.&(1.11)\cr}$$

\nt where $N_n=(m\w/\pi\hbar)^{1/4}/(2^n n!)^{1/2}$. From the parametrised
Lagrangian we find the conserved quantity

$$Q={m\o2}\,[{\dot x}^2+2\,\G\,{\dot x}\,x+{\w_0}^2\,x^2]\,e^{2\G t}={m\o2}\,
[({\dot x}+\G\,x)^2+\w^2\,x^2]\,e^{2\G t}\,.\eq{(1.12)}$$

\nt It is easy to check that the wavefunctions (1.11) are eigenfunctions of
${\hat Q}$

$${\hat Q}\,\psi=q\,\psi\,,\eq{(1.13)}$$

\nt where the eigenvalue $q$ is given by

$$q=\left(n+{1\o2}\right)\,\hbar\,\w\,.\eq{(1.14)}$$

Therefore, the procedure of parametrisation and use of the Noether theorem
allows the construction of conserved quantities which can be used for the
construction of pure states.

In section 2 we present standard results for systems described by
time-independent Lagrangians. In section 3 we show how all the above nice
prescriptions fail for a time-dependent system. Section 4 illustrates how
parametrised mechanics works. Section 5 is devoted to the example. The
conclusions and comments on future developments are contained in section 6.

\bs\bs
\nt{\bf 2. Time-Independent Systems}
\bs

Here we present the general results for mechanics, conservation laws and
quantum mechanics for systems described by a time-independent Lagrangian.

\bs\bs
\nt{\bf 2.1. Classical Mechanics}
\bs

We will restrict our considerations to mechanical systems described by a
Lagrangian

$$L=L(q,\,\dq)\,.\eq{(2.1)}$$

\nt The equations of motion are given by

$${{\d L}\o{\d q^i}}={{\p L}\o{\p q^i}}-{d\o{dt}}\left({{\p L}\o{\p\dq^i}}
\right)=0\,.\eq{(2.2)}$$

\nt Contraction of the equations of motion with $\dq^i$ gives

$${{d E}\o{d t}}=\dq^i\,{{\d L}\o{\d q^i}}=0\,,\eq{(2.3)}$$

\nt where $E$ is the energy of the system

$$E=\dq^i\,{{\p L}\o{\p\dq^i}}-L\,.\eq{(2.4)}$$

The canonical momenta are defined by

$$p_i={{\p L}\o{\p\dq^i}}\,,\eq{(2.5)}$$

\nt and the canonical Hamiltonian is given by

$$H=\dq^i\,p_i-L\,.\eq{(2.6)}$$

\nt The Hamiltonian is a function of $q$'s and $p$'s only, $H=H(q,p)$, and
furthermore is a conserved quantity. The equations of motion (2.2) are
replaced by the Hamilton equations

$$\eqalignno{\dq^i&={{\p H}\o{\p p_i}}\,,\cr
{\dot p}_i&=-{{\p H}\o{\p q^i}}\,,&(2.7)\cr}$$

\nt The time derivative of a phase space function $F=F(q,p)$ is given by

$${\dot F}=\{F,\,H\}\,,\eq{(2.8)}$$

\nt where $\{,\}$ is the Poisson bracket defined by

$$\{F,\,G\}={{\p F}\o{\p q^i}}\,{{\p G}\o{\p p_i}}-{{\p G}\o{\p q^i}}\,{{\p F
}\o{\p p_i}}\,.\eq{(2.9)}$$

\bs\bs
\nt{\bf 2.2. The Noether theorem}
\bs

The most systematic way for finding conserved quantities is through the
Noether theorem. Let us define the Noether variation of the Lagrangian

$$\xi(L)=\xi^i\,{{\p L}\o{\p q^i}}+{\dot\xi}^i\,{{\p L}\o{\p\dq^i}}\,.
\eq{(2.10)}$$

\nt We furthermore define the charge

$$Q(\xi)=\xi^i\,{{\p L}\o{\p\dq^i}}=\xi^i\,p_i\,.\eq{(2.11)}$$

\nt The time derivative of $Q(\xi)$ is given by

$${\dot Q}(\xi)=\xi(L)-\xi^i\,{{\d L}\o{\d q^i}}\,.\eq{(2.12)}$$

\nt Therefore, if the equations of motion (2.2) hold and if

$$\xi(L)=0\,,\eq{(2.13)}$$

\nt then $Q(\xi)$ is a conserved quantity.

\bs\bs
\nt{\bf 2.3. Quantum Mechanics}
\bs

For our purposes we can consider quantum mechanics just as an operator
realisation of the Poisson bracket algebra. In the coordinate representation

$$\eqalignno{q^i\,&\ra\,q^i\,,\cr
p_i\,&\ra\,-\im\,\hbar\,{\p\o{\p q^i}}\,.&(2.14)\cr}$$

\nt Furthermore, all classical functions are transformed in differential
operators,

$$F\,\ra\,{\hat F}=F({\hat q},\,{\hat p})\,,\eq{(2.15)}$$

\nt with the additional restriction

$$[{\hat F},\,{\hat G}]=\im\,\hbar\,{\widehat{\{F,\,G\}}}\,.\eq{(2.16)}$$

There is furthermore a wave function $\psi=\psi(q,t)$ satisfying the
Schr\"odinger equation

$${\hat H}\,\psi=H\left(q,\,-\im\,\hbar\,{\p\o{\p q}}\right)\,\psi=\im\,\hbar
\,{{\p\psi}\o{\p t}}\,.\eq{(2.17)}$$

\nt Since the Hamiltonian is time independent we can write

$$\psi(q,\,t)=\vf(q)\,e^{-\im\w t}\,,\eq{(2.18)}$$

\nt and the Schr\"odinger equation reduces to

$${\hat H}\,\vf=H\left(q,\,-\im\,\hbar\,{\p\o{\p q}}\right)\,\vf=\hbar\,\w\,
\vf=E\,\vf\,.\eq{(2.19)}$$

\nt The above is nothing more than the quantum mechanical manifestation of
the classical conservation law for the energy, (2.4). The solutions to eq.
(2.19) define pure states, {\it i.e.} states with a well defined quantum
number with respect to a given observable, the Hamiltonian in this case.

\bs\bs
\nt{\bf 3. Time-Dependent Systems}
\bs

We now present the classical mechanics, conservation laws and quantum
mechanics for systems described by time-dependent Lagrangians.

\bs\bs
\nt{\bf 3.1. Classical Mechanics}
\bs

Let us now consider a mechanical systems described by a time-dependent
Lagrangian

$$L=L(q,\,\dq,\,t)\,.\eq{(3.1)}$$

\nt The equations of motion are given, as before, by (2.2). The energy $E$ is
defined as in (2.4). Now, however, the time derivative of the energy is

$${{d E}\o{d t}}=\dq^i\,{{\d L}\o{\d q^i}}-{{\p L}\o{\p t}}\not=0\,.
\eq{(3.2)}$$

The Hamiltonian formalism remains almost unchanged. The momenta are defined
as in (2.5) and the Hamiltonian as in (2.6), however, now the Hamiltonian is
a function also of $t$, $H=H(q,p,t)$. The Hamilton equations (2.7) are
unchanged. This time, however, the time derivative of a phase space function
$F=F(q,p,t)$ is given by

$${\dot F}={{\p F}\o{\p t}}+\{F,\,H\}\,,\eq{(3.3)}$$

\nt where $\{,\}$ is the Poisson bracket defined in (2.9).

The results of Section 2.2, concerning the Noether theorem, remain unchanged
for time-dependent systems. However, the Noether theorem does not provide
useful conserved quantities to be used for quantisation.

\bs\bs
\nt{\bf 3.2. Quantum Mechanics}
\bs

Once again we can consider quantum mechanics just as an operator realisation
of the Poisson bracket. In the coordinate representation eqs. (2.10) are
still valid, and all classical functions are transformed in differential
operators, as in (2.11) with the additional restriction (2.12).

There is furthermore a wave function $\psi=\psi(q,t)$ satisfying the
Schr\"odinger equation

$${\hat H}\,\psi=H\left(q,\,-\im\,\hbar\,{\p\o{\p q}},\,t\right)\,\psi=\im\,
\hbar\,{{\p\psi}\o{\p t}}\,.\eq{(3.4)}$$

\nt Now, however, the simple prescription (2.18) does not work anymore. The
solutions will be, therefore, unavoidably of the form $\psi=\psi(q,t)$. In
this case it is no more possible to talk of stationary states. Furthermore
the wavefunction is not a pure state with respect to the Hamiltonian.
Finally, it is not clear if there exist other observables with respect to
which the wavefunction is a pure state.

The problem is therefore reduced to the construction of conserved quantities,
if any, with respect to which the wave function $\psi(q,t)$ is a pure state.
The solution to this problem is provided by parametrised mechanics.

\bs\bs
\nt{\bf 4. Parametrised Mechanics}
\bs

Here we present parametrised classical mechanics, conservations laws, and
quantum mechanics.

\bs\bs
\nt{\bf 4.1. Parametrising unparametrised mechanics}
\bs

In parametrised mechanics the time $t$ is put at the same level than the
configuration variables $q$, $t\ra q^0$. In this case the role of the time is
played by a new parameter $\t$. The new Lagrangian is obtained by requiring
invariance of the action

$$S=\int\,L(q,\,{\bar q},\,t)\,dt=\int\,L(q,\,\q,\,q^0)\,\dq^0\,d\t\,,
\eq{(4.1)}$$

\nt where now derivatives with respect to the old time are denoted by a bar,
and dots denote derivatives with respect to $\t$; furthermore

$$\q^i={{\dq^i}\o{\dq^0}}\,.\eq{(4.2)}$$

\nt Therefore, the new Lagrangian is given by

$${\bar L}=L(q,\,\q,\,q^0)\,\dq^0\,.\eq{(4.3)}$$

\nt The equations of motion are given by

$$\eqalignno{{{\d{\bar L}}\o{\d q^i}}&={{\p{\bar L}}\o{\p q^i}}-{d\o{d\t}}
\left({{\p{\bar L}}\o{\p\dq^i}}\right)={{\p L}\o{\p q^i}}\,\dq^0-{d\o{d\t}}
\left({{\p L}\o{\p\q^i}}\right)=0\,,\cr
{{\d{\bar L}}\o{\d q^0}}&={{\p{\bar L}}\o{\p q^0}}-{d\o{d\t}}\left({{\p{\bar
L}}\o{\p\dq^0}}\right)={{\p L}\o{\p q^0}}\,\dq^0-{d\o{d\t}}\left(-\q^i\,{{\p
L}\o{\p\q^i}}+L\right)=0\,.&(4.4)\cr}$$

The momenta are given by

$$\eqalignno{{\bar p}_0&={{\p{\bar L}}\o{\p\dq^0}}=-{{\p L}\o{\p\q^i}}\,\q^i+
L=-H(q^i,\,q^0,\,\q^i)\,,&(4.5a)\cr
{\bar p}_i&={{\p{\bar L}}\o{\p\dq^i}}={{\p L}\o{\p\q^i}}\,,&(4.5b)\cr}$$

\nt The Hamiltonian is given by

$${\bar H}=\dq^0\,{\bar p}_0+\dq^i\,{\bar p}_i-{\bar L}\equiv0\,.\eq{(4.6)}$$

If the original Lagrangian was regular then, such as $\dq$'s are solved in
terms of $p$'s, eq. (4.5b) can be solved for $\q$'s in terms of ${\bar p}$'s.
Therefore (4.5a) can be rewritten as

$${\bar p}_0=-H(q^i,\,q^0,\,{\bar p}_i)\,.\eq{(4.7)}$$

\nt The above relation is the primary first-class constraint associated to
the parametrisation invariance of the Lagrangian [1]

$$\f={\bar p}_0+H(q^i,\,q^0,\,{\bar p}_i)\approx0\,.\eq{(4.8)}$$

As is well known, for constrained systems quantum mechanics is implemented in
terms of first-class constraints. In the configuration representation

$$\eqalignno{q^i\,&\ra\,q^i\,,\cr
q^0\,&\ra\,t\,,\cr
{\bar p}_i\,&\ra\,-\im\,\hbar\,{\p\o{\p q^i}}\,,\cr
{\bar p}_0\,&\ra\,-\im\,\hbar\,{\p\o{\p t}}\,.&(4.9)\cr}$$

\nt For the first-class constraint we obtain

$${\hat\f}=-\im\,\hbar\,{\p\o{\p t}}+H\left(q^i,\,t,\,-\im\,\hbar\,{\p\o{\p q
}}\right)\approx0\,.\eq{(4.10)}$$

\nt The physical states $\psi$ are those annihilated by (4.10)

$${\hat\f}\,\psi=-\im\,\hbar\,{{\p\psi}\o{\p t}}+H\left(q^i,\,t,\,-\im\,\hbar
\,{\p\o{\p q^i}}\right)\,\psi=0\,,\eq{(4.11)}$$

\nt which we recognise as the original Schr\"odinger equation.

Therefore, the parametrised theory contains the standard results. However,
this does not provide a solution to our problem: the search for conserved
quantities.

\bs\bs
\nt{\bf 4.2. The Noether theorem for parametrised mechanics}
\bs

The Noether theorem applies unchanged to parametrised mechanics. In this case
however, the configuration space is extended by the addition of the old time
variable as a configuration variable. Therefore, it is convenient to write
down explicitly all the corresponding formulae for this case.

In this case the Noether variation of the Lagrangian is given by

$$\eqalignno{\xi({\bar L})&=\xi^0\,{{\p{\bar L}}\o{\p q^0}}+\xi^i\,{{\p{\bar
L}}\o{\p q^i}}+{\dot\xi}^0\,{{\p{\bar L}}\o{\p\dq^0}}+{\dot\xi}^i\,{{\p{\bar
L}}\o{\p\dq^i}}\cr
&=\xi^0\,{{\p L}\o{\p q^0}}\,\dq^0+\xi^i\,{{\p L}\o{\p q^i}}\,\dq^0+{\dot\xi}
^0\,\left(-\q^i\,{{\p L}\o{\p\q^i}}+L\right)+{\dot\xi}^i\,{{\p L}\o{\p\q^i}}
&(4.12)\cr}$$

\nt The Noether charge is defined as

$$Q(\xi)=\xi^0\,{{\p{\bar L}}\o{\p\dq^0}}+\xi^i\,{{\p{\bar L}}\o{\p\dq^i}}=
\xi^0\,{\bar p}_0+\xi^i\,{\bar p}_i\,.\eq{(4.13)}$$

\nt Once again, if the equations of motion hold, and if

$$\xi({\bar L})=0\,,\eq{(4.14)}$$

\nt then the Noether charge $Q(\xi)$ is a conserved quantity

$${{dQ}\o{d\t}}=0\,.\eq{(4.15)}$$

\nt It is also a conserved quantity with respect to the original time

$${{dQ}\o{dt}}={{d\t}\o{dt}}\,{{dQ}\o{d\t}}=0\,.\eq{(4.16)}$$

By using the constraint (4.8) we observe that $Q$ can be written as

$$Q(q^i,\,q^0,\,{\bar p}_i)=-\xi^0(q^i,\,q^0)\,H(q^i,\,q^0,\,{\bar p}_i)+\xi^
i(q^i,\,q^0)\,{\bar p}_i\,.\eq{(4.17)}$$

\nt Therefore it is something more than the Hamiltonian alone. The corrective
term $\xi^i\,{\bar p}_i$ is able to kill the effect of the time dependence
and give a conserved quantity.

\bs\bs
\nt{\bf 4.3. Parametrised Quantum Mechanics}
\bs

Quantum mechanics is performed in terms of the quantum version of $Q$, namely

$${\hat Q}=Q({\hat q}^i,\,{\hat q}^0,\,{\hat p}_i)=-\xi^0({\hat q}^i,\,{\hat
q}^0)\,H({\hat q}^i,\,{\hat q}^0,\,{\hat p}_i)+\xi^i({\hat q}^i,\,{\hat q}^0)
\,{\hat p}_i\,.\eq{(4.18)}$$

\nt Of course, in (4.18) may appear operator ordering problems. The
corresponding ``Schr\"odin-ger'' equation is

$${\hat Q}\,\psi=\left[-\xi^0({\hat q}^i,\,{\hat q}^0)\,H({\hat q}^i,\,{\hat
q}^0,\,{\hat p}_i)+\xi^i({\hat q}^i,\,{\hat q}^0)\,{\hat p}_i\right]\,\psi=q
\,\psi\,.\eq{(4.19)}$$

Therefore, the Schr\"odinger equation is modified to an equation for pure
states.

\bs\bs
\nt{\bf 5. The Damped Harmonic Oscillator}
\bs

The damped harmonic oscillator is the simplest system described by a
time-dependent Lagrangian. Some of the results presented here are standard.
We refer the reader to [2] for further details.

\bs\bs
\nt{\bf 5.1. Classical mechanics of the damped harmonic oscillator}
\bs

The damped harmonic oscillator is described by the equation

$${\ddot x}+2\,\G\,{\dot x}+{\w_0}^2\,x=0\,,\eq{(5.1)}$$

\nt where $\G>0$. The usual Lagrangian used to describe the damped harmonic
oscillator is the Bateman Lagrangian [3]

$$L={m\o2}\,({\dot x}^2-{\w_0}^2\,x^2)\,e^{2\G t}\,.\eq{(5.2)}$$

\nt The energy would be

$$E={m\o2}\,({\dot x}^2+{\w_0}^2\,x^2)\,e^{2\G t}\,.\eq{(5.3)}$$

The momentum is given by

$$p=m\,{\dot x}\,e^{2\G t}\,.\eq{(5.4)}$$

\nt The Hamiltonian is then

$$H={{p^2}\o{2m}}\,e^{-2\G t}+{m\o2}\,{\w_0}^2\,x^2\,e^{2\G t}\,.\eq{(5.5)}$$

\bs\bs
\nt{\bf 5.2. Quantum mechanics of the damped harmonic oscillator}
\bs

The Schr\"odinger equation associated to the Hamiltonian (5.5) is

$$\left[-{{\hbar^2}\o{2m}}\,e^{-2\G t}\,{{\p^2}\o{\p x^2}}+{{m{\w_0}^2}\o2}\,
e^{2\G t}\,x^2\right]\,\psi=\im\,\hbar\,{{\p\psi}\o{\p t}}\,.\eq{(5.6)}$$

\nt Due to the explicit appearance of time in the Hamiltonian, this equation
does not admit stationary solutions. The solutions to eq. (5.6) is given by

$$\eqalignno{\psi_n(x,\,t)&=N_n\,H_n\left(\left({{m\w}\o\hbar}\right)^{1/2}\,
e^{\G t}\,x\right)\cr
&\quad\times\exp\left(-{m\o{2\hbar}}(\w+\im\G)e^{2\G t}x^2\right)\,\exp\left(
-\im\left(n+{1\o2}\right)\w t+{\G\o2}\,t\right)\,.&(5.7)\cr}$$

\nt where $N_n$ is a normalisation constant given by

$$N_n={1\o{(2^n n!)^{1/2}}}\,\left({{m\w}\o{\pi\hbar}}\right)^{1/4}\,,
\eq{(5.8)}$$

\nt and

$$\w^2={\w_0}^2-\G^2>0\,,\eq{(5.9)}$$

The wavefunction depends explicitly on time, therefore it is not stationary.
Furthermore it is not a pure state.

\bs\bs
\nt{\bf 5.3. The parametrised damped harmonic oscillator}
\bs

The parametrised damped harmonic oscillator is described by the Lagrangian

$$L={m\o2}\,\left({{{\dot x}^2}\o{\dot\t}}-{\w_0}^2\,x^2\,{\dot\t}\right)\,e^
{2\G t}\,.\eq{(5.10)}$$

\nt Using the Noether theorem as applied to parametrised mechanics (Sec. 4.2)
we find the following conserved quantity

$$Q={m\o2}\,[{\dot x}^2+2\,\G\,{\dot x}\,x+{\w_0}^2\,x^2]\,e^{2\G t}={m\o2}\,
[({\dot x}+\G\,x)^2+\w^2\,x^2]\,e^{2\G t}\,.\eq{(5.11)}$$

\nt In fact, the time derivative of this quantity is

$${\dot Q}=m\,({\dot x}+\G\,x)\,({\ddot x}+2\,\G\,{\dot x}+{\w_0}^2\,x)\,e^{2
\G t}=0\,,\eq{(5.12)}$$

\nt which is zero in virtue of the equations of motion (5.1). Let us finally
observe that, for $\G=0$, $Q$ reduces to the usual energy.

Up to our knowledge the above quantity has not been reported as a conserved
quantity associated to the damped harmonic oscillator.

\bs\bs
\nt{\bf 5.4. Quantum mechanics of the parametrised damped harmonic
oscillator}
\bs

Now we use $Q$ for doing quantum mechanics. The corresponding operator is
given by

$${\hat Q}=-{{\hbar^2}\o{2m}}\,e^{-2\G t}\,{{\p^2}\o{\p x^2}}-\im\,\hbar\,\G
\,\left(x\,{\p\o{\p x}}+{1\o2}\right)+{m\o2}\,{\w_0}^2\,x^2\,e^{2\G t}\,.
\eq{(5.13)}$$

We can next construct the Schr\"odinger-like equation

$${\hat Q}\,\psi=\left[-{{\hbar^2}\o{2m}}\,e^{-2\G t}\,{{\p^2}\o{\p x^2}}-\im
\,\hbar\,\G\,\left(x\,{\p\o{\p x}}+{1\o2}\right)+{m\o2}\,{\w_0}^2\,x^2\,e^{2
\G t}\right]\,\psi=q\,\psi\,.\eq{(5.14)}$$

\nt The factor ${1\o2}$ in the middle term appears after operator ordering,
cf. [4]. We can verify that the solution is given exactly by (5.7). The
eigenvalue $q$ is given by

$$q=\left(n+{1\o2}\right)\,\hbar\,\w\,.\eq{(5.15)}$$

In conclusion, $Q$ is the observable with respect to which the wave function
$\psi$, solution of the time dependent Schr\"odinger equation, behaves as a
pure state.

\bs\bs
\nt{\bf 6. Conclusions}
\bs

We have introduced a method which allows to interpret the time-dependent
wavefunctions as pure states with respect to another observables. The method
is based on the construction of conserved quantities associated to
time-dependent systems. The method itself deserves more consideration since
it provides new conserved quantities for time-dependent systems. In fact,
(5.11) is a conserved quantity for the damped harmonic oscillator. However,
it caannot be found by direct application of the Noether theorem to the
unparametrised Lagrangian (5.2). This is an example of a non-Noetherian
conserved quantity.

\bs\bs
\nt{\bf Acknowledgements}
\bs

This work has been partially supported by Direcci\'on de Investigaci\'on,
Universidad de Concepci\'on, under contract 94.11.08-1. The work took its
final form thanks to several discussions with J. C. Retamal and C. Saavedra.

\bs\bs
\nt{\bf References}
\bs

\item{[1]}Dirac P A M 1964 {\it Lectures on Quantum Mechanics} Belfer
Graduate School of Science (Yeshiva University, New York)

\item{[2]}Dekker H 1980 {\it Phys. Rep.} {\bf 80} 1

\item{[3]}Bateman H 1931 {\it Phys. Rev.} {\bf 38} 815

\item{[4]}Christodoulakis T, and Zanelli J 1986 {\it Nuovo Cimento} B {\bf
93} 1

\bye